\documentclass[twocolumn]{aastex631}

\usepackage{amsmath} 
\usepackage{amssymb}
\usepackage{graphicx}
\usepackage{wrapfig}
\graphicspath{{images/}}  
\usepackage{multirow} 
\usepackage{booktabs}
\usepackage{hyperref}
\usepackage{nameref} 
\usepackage{natbib} 
\begin{document}

\title{\Large Distance to the Galactic centre 7$^\prime$ halo from HI absorption}

\author[0009-0003-6833-7048]{Debangan Maji} 
\affiliation{National Centre for Radio Astrophysics - Tata Institute of Fundamental Research, \\ Pune University Campus, Post Bag 3, Ganeshkhind Pune 411007, INDIA
\\ email: dmaji@ncra.tifr.res.in, roy@ncra.tifr.res.in}

\author{Subhashis Roy}
\affiliation{National Centre for Radio Astrophysics - Tata Institute of Fundamental Research, \\ Pune University Campus, Post Bag 3, Ganeshkhind Pune 411007, INDIA
\\ email: dmaji@ncra.tifr.res.in, roy@ncra.tifr.res.in}

\begin{abstract}
    Using archival data, we have made an HI absorption study of the 7$^\prime$ halo surrounding the Sgr A complex, observed towards the Galactic centre (GC) region. We find strong HI absorption near velocities of \ensuremath{-53} km s\ensuremath{^{-1}}, which is due to the 3-kpc arm, placing it beyond 5 kpc from us. We further examined the HI absorption properties towards 5 different parts of the 7$^\prime$ halo. Absorption by \ensuremath{+50} km s\ensuremath{^{-1}} GC cloud is seen towards only 3 parts of the halo, but not towards the other 2 regions. Strong emissions in CO and CS are, however, identified toward all the above 5 parts of the halo by the \ensuremath{+50} km s\ensuremath{^{-1}} GC molecular cloud. This does show that the 7$^\prime$ halo is partly behind, and partly in front of the \ensuremath{50} km s\ensuremath{^{-1}} cloud. To our knowledge, this, for the first time clearly shows the 7$^\prime$ halo to be located at the same distance as the \ensuremath{+50} km s\ensuremath{^{-1}} molecular cloud, i.e., at the GC region.
\end{abstract}


\section{Introduction}
\label{sec:introduction}

The central one-kilo-parsec region of our Milky Way (MW) Galaxy is typically referred to as the Galactic centre (GC) region. Most physical quantities like gas density, velocity dispersion, temperature, etc. in this region are significantly higher than those found in the disk of our galaxy \citep{Bally1987, Morris1996}. Near the dynamic centre of MW, there is a moderately powerful compact radio source called Sgr $\text{A}^*$, which coincides with the black hole of mass $4.28 \times 10^6 M_\odot$ \citep{Gillessen2017}. The region around Sgr $\text{A}^*$ shows emission at varieties of size scales. Around the Sgr $\text{A}^*$ region is the three armed spiral structure of the HII region Sgr A West. Sgr A West is encompassed by the supernova remnant Sgr A East ($\sim3^\prime$), and around Sgr A East is seen the diffuse emission of 7$^{\prime}$ halo. The ensemble of the above 4 objects is generally inferred as Sgr A complex region (see Fig. 5 of \citet{Anantharamaiah1991} for a schematic drawing of the region). 

Existing continuum, HI, and molecular line emission studies have indicated that Sgr $\text{A}^*$, Sgr A West and East are all located very close to the GC, and are not a case of chance superpositions along our line of sight. Free-free absorption study towards the Sgr A complex has shown that Sgr A West is located in front of Sgr A East. \cite{pedlar1989sgrA} claimed that the 7$^\prime$ halo is a mixture of thermal and non-thermal emission because they found that the radio spectrum has low-frequency turnover and is flatter than expected for just synchrotron emission. They inferred that self-absorption, because of the thermal component, is responsible for the low-frequency turnover. However, based on 150 MHz observation, \cite{RoyandRao2006} showed that the absorption spectrum of 7$^\prime$ halo and Sgr A East appear very similar to each other below 0.4 GHz (Figs. 5 \& 6 in \cite{RoyandRao2006}). They ascribed the low frequency absorption to a foreground ionized gas, and then the 7$^\prime$ halo is mostly a synchrotron-emitting nebulae.

Interferometric HI absorption studies of the region \citep{Radhakrishnan1972,Schwarz1982,Liszt1983,Dwarakanath2004} identified a wide HI line towards Sgr $\text{A}^*$ and suggested the circumnuclear disk (CND) to be responsible for it. HI absorption studies towards Sgr A West \& East were carried out by \cite{Lang2010}. However, to our knowledge, no limit on the line of sight distance of 7$^\prime$ halo has been placed in the literature. A lack of constraints on its distance limits our understanding of the origin of the nebulae. We have imaged this halo in HI absorption using VLA archival data (VLA observation AL546). Here we present the results and constrain its line of sight distance.

\section{Data Reduction \& Analysis}

We used VLA archival HI data from the project AL546. Astronomical Image Processing System (AIPS) and Common Astronomy Software Applications (CASA) were used to analyse the data following a standard approach. The survey contains five different days of observations. For calibration, we analysed the data for each day separately. We have flagged (rejected) bad outlier data based on their deviation from smoothness in time and frequency. Both time and frequency-based calibrations were done using standard calibrator sources (3C286, 3C48 and 1751$-$253). After calibrating each of the five data sets, we combined them, and then made continuum images for each of the fields in the data. The continuum sources were subtracted from the UV data using a clean component based subtraction (UVSUB in AIPS), and then fitted for a DC-term from the line-free frequency channels (channels $8-15$ and $103-120$), which was then subtracted from the input UV data. The continuum subtracted data sets contained 5 fields, which were all separated by half of the primary beam-width of the antennas. For improved sensitivity, we made a mosaic image (HI absorption cube) of all the fields using the CASA task TCLEAN. A mosaic of a continuum image at 21cm using line-free channels was also made, and the image rms is about 5 mJy/beam. Primary beam correction was performed for all the images before further analysis.

The observations were taken before the VLA WIDAR correlator upgrade, where the Doppler tracking method was used as default for VLA observations (see \href{https://science.nrao.edu/facilities/vla/docs/manuals/test-manual/line}{NRAO VLA Spectral Line Observing Manual}\footnote{\url{https://science.nrao.edu/facilities/vla/docs/manuals/test-manual/line}}), and therefore we have not applied any Doppler correction to our data. An HI survey from this data set has been published earlier in \cite{Lang2010}.

To further investigate the GC region, we have also used CO and CS molecular emission survey cubes by \citet{Tokuyama2019} and \citet{Tsuboi1999}, respectively. We have convolved every cube to the same resolution. These emission line cubes were used to make the spectra of 7$^\prime$ halo regions, which are discussed in section~\ref{sec:results}. We have also made the CS emission contour maps overlap with the 1.4 GHz continuum emission of the 7$^\prime$ halo region which was done using AIPS task KNTR and further discussed in section~\ref{sec:discussions}.

\section{Results}
\label{sec:results}
\begin{figure}[h!]
    \centering
    \includegraphics[width=1\linewidth]{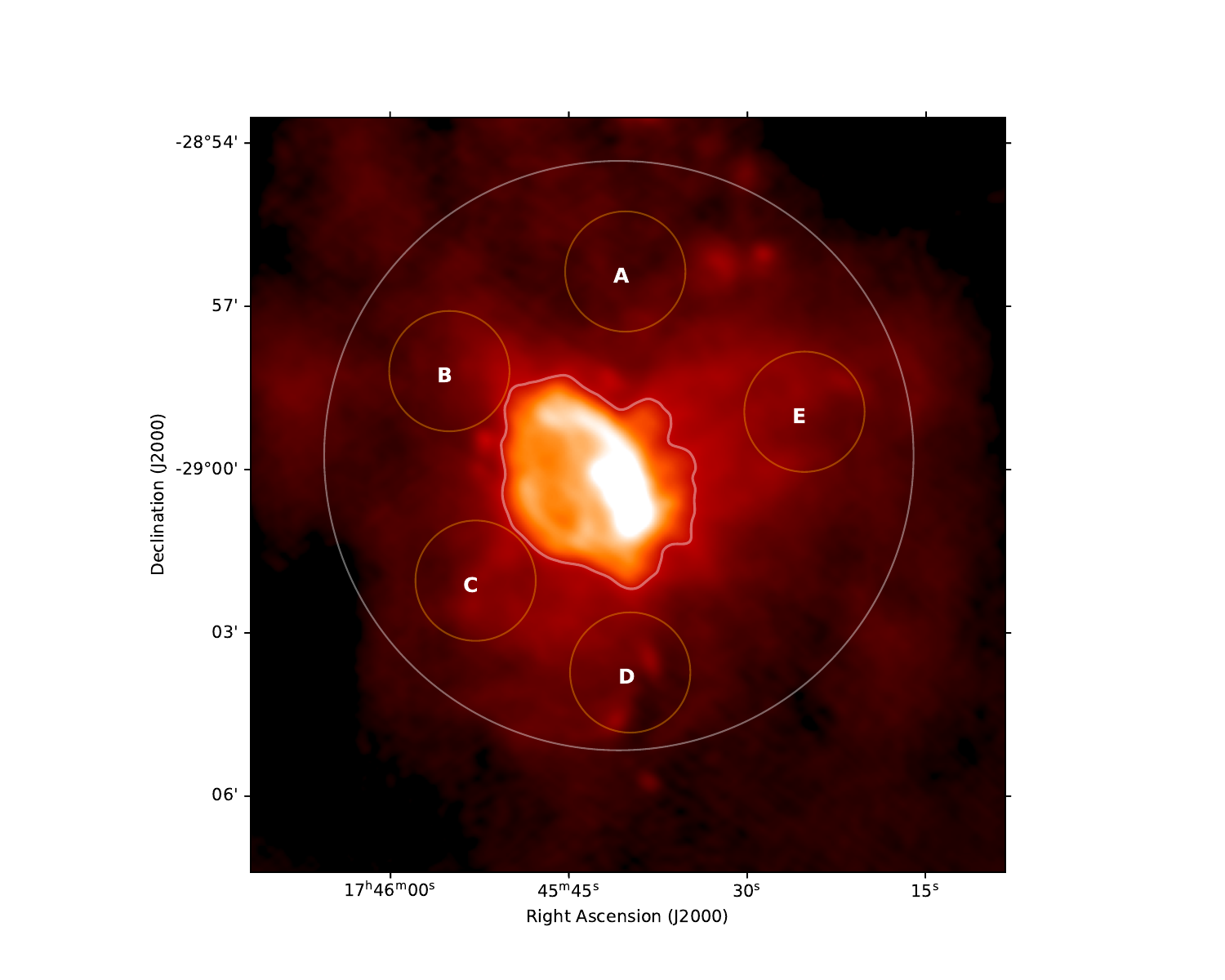}
    \caption{Continuum image of the 7$^\prime$ halo at 1.42 GHz}
    \label{fig:7-arcmin halo}
\end{figure}

\begin{figure}[h!]
    \centering
    \includegraphics[width=1\linewidth]{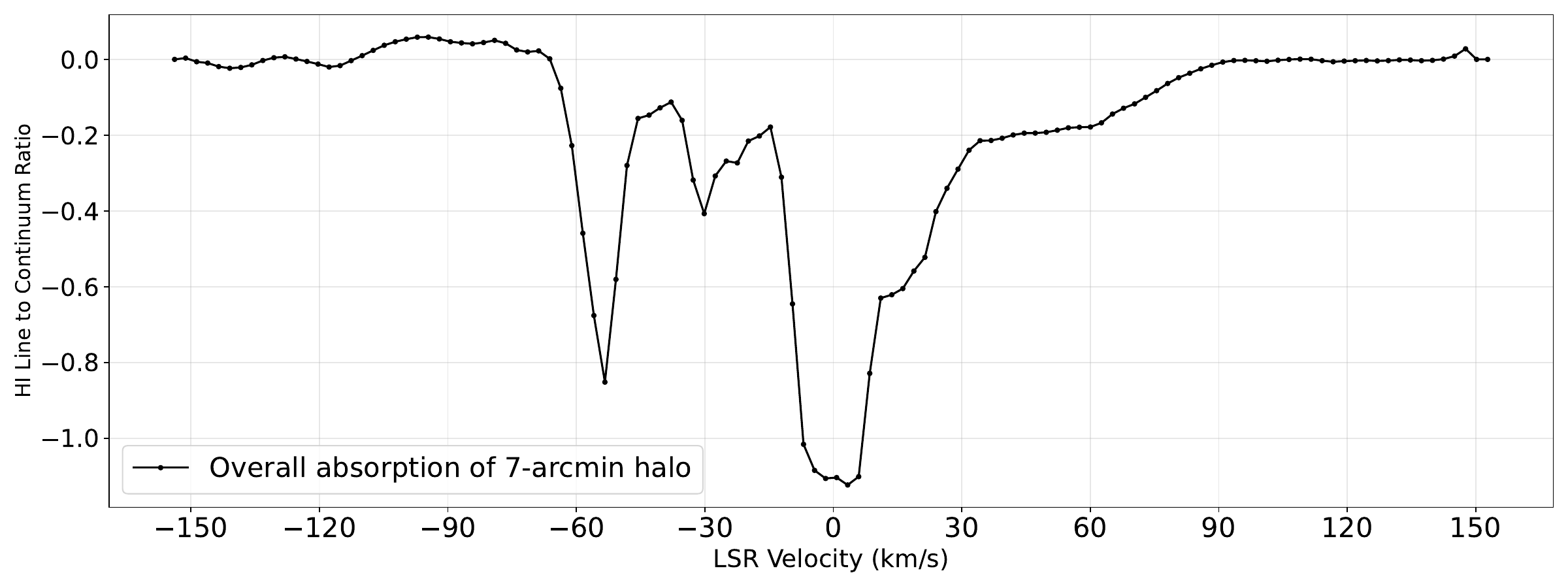}
    \caption{HI absorption of the 7$^\prime$ halo}
    \label{fig: HI absorption of 7-arcmin halo}
\end{figure}

\begin{figure*}
    \centering
    \includegraphics[width=1\linewidth]{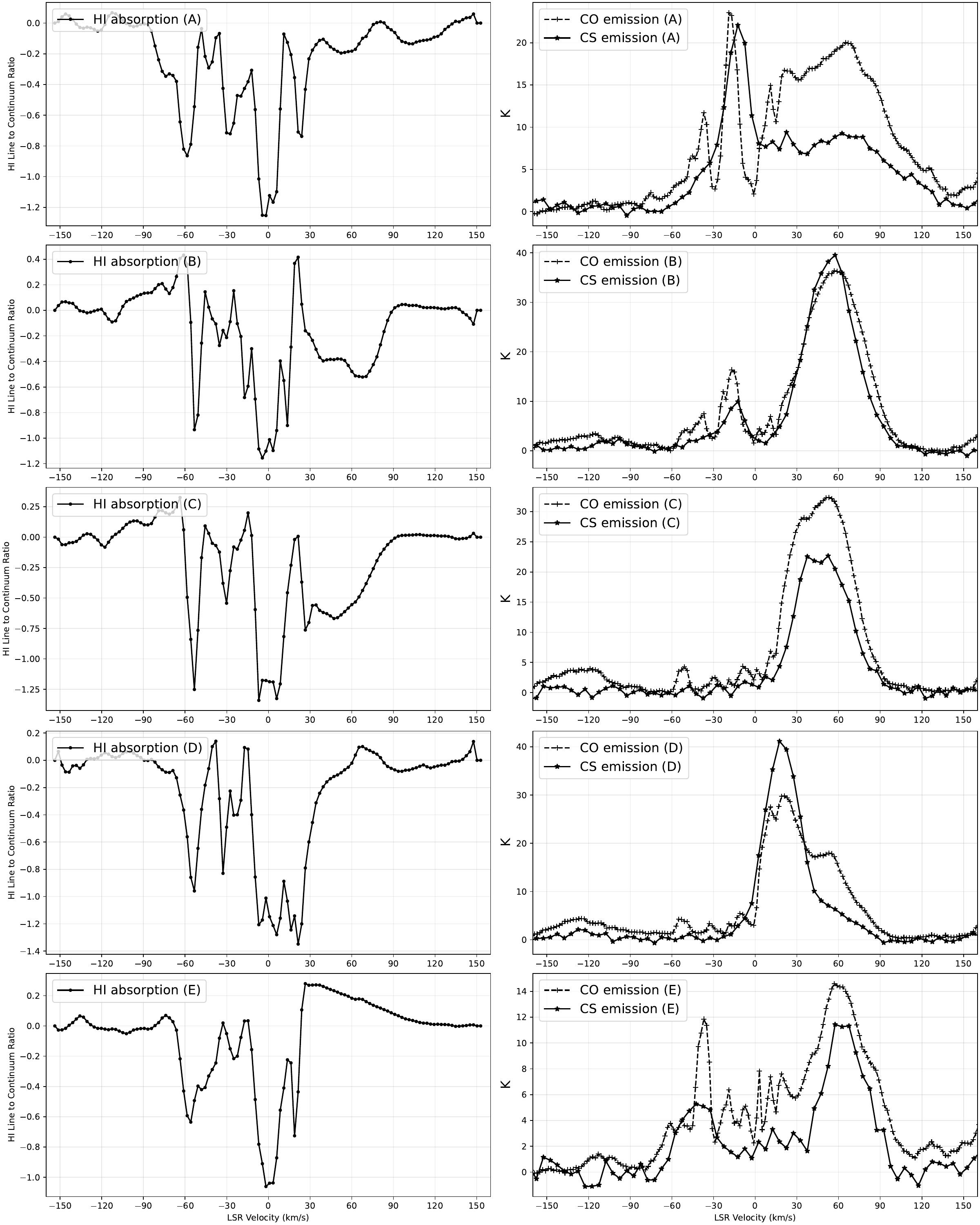}
    \caption{(a) Left column shows the HI absorption for the different parts (Figure~\ref{fig:7-arcmin halo}) of the 7$^\prime$ halo (b) right column shows the CO \citep{Tokuyama2019} and CS \citep{Tsuboi1999} molecular emission from the same parts, CS emission values are scaled by a factor of 20 to overlap the plots with CO emission.}
    \label{fig:hi_co_cs}
\end{figure*}

The continuum image of Sgr A complex is shown in Figure~\ref{fig:7-arcmin halo}. To find the spectrum of the 7$^\prime$ halo, we have integrated over the annular region between the large circular contour and the contour surrounding Sgr A East, which are also shown in Figure~\ref{fig:7-arcmin halo}. Figure~\ref{fig: HI absorption of 7-arcmin halo} shows the integrated spectrum. A clear absorption feature is seen at a velocity of about $-53$ km s$^{-1}$. This absorption feature coincides with the known emission velocity of the 3-kpc arm \citep{Rougoor1964}. In addition, absorption is also seen near $-$30 km s$^{-1}$. Broad absorption features could also be seen near +50 km s$^{-1}$. To investigate the absorption features across the 7$^\prime$ halo with more details, we have also generated spectra from 5 different regions of the halo. These regions are shown with yellow circles and marked with letters A to E in Figure~\ref{fig:7-arcmin halo}, and the corresponding absorption features are shown in Figure~\ref{fig:hi_co_cs}. The velocities provided for each plot are measured in the Local Standard of Rest (LSR) reference frame.

Table~\ref{tab:absorption information} shows the LSR velocity and the corresponding optical depth of HI absorption seen toward different regions of the 7$^\prime$ halo. As mentioned earlier, absorption near $-$30 km s$^{-1}$ is seen in the integrated spectrum (Figure~\ref{fig: HI absorption of 7-arcmin halo}), and the velocities of the absorption peak of this feature vary from $-25$ (for regions D and E) to $-$30 km s$^{-1}$ (for regions B and C). In addition, we find a broad HI absorption at positive velocities from 50 km s$^{-1}$ (regions A , C) to about 68 km s$^{-1}$ (region B). 

\begin{table}[h]
    \centering
    \caption{LSR velocity and corresponding Optical Depth of HI absorption seen in different locations}   
    \resizebox{\columnwidth}{!}{
    \begin{tabular}{c c c c}
        \toprule
        Parts of & \multicolumn{1}{c}{Velocity} & Optical Depth & Notes \\
        $7^\prime$ halo & \multicolumn{1}{c}{(km s$^{-1}$)} & ($\tau$) & \\
        \midrule
        \multirow{7}{*}{A}  
        & $-$73.9  & 0.4$\pm$0.1  & -\\
        & $-$58.5  & 2.0$\pm$0.2  & 3-kpc arm \\
        & $-$43.0  & 0.4$\pm$0.1  & -\\
        & $-$27.6  & 1.3$\pm$0.1  & -\\
        & +24.0  & 1.3$\pm$0.1  & -\\
        & +52.3  & 0.2$\pm$0.1  & +50 km s$^{-1}$ \\ & & & molecular cloud\\
        & +103.8 & 0.2$\pm$0.1  & -\\
        \midrule
        \multirow{8}{*}{B}  
        & $-$112.6 & 0.1$\pm$0.1  & -\\
        & $-$53.3  & 2.7$\pm$0.2  & 3-kpc arm \\
        & $-$35.3  & 0.3$\pm$0.1  & -\\
        & $-$30.1  & 0.2$\pm$0.1  & -\\
        & $-$17.3  & 1.2$\pm$0.1  & -\\
        & +13.7  & 2.3$\pm$0.3  & -\\
        & +39.4  & 0.5$\pm$0.1  & +50 km s$^{-1}$ \\ & & & molecular cloud\\
        & +67.8  & 0.7$\pm$0.1  & -\\
        \midrule
        \multirow{4}{*}{C}  
        & $-$53.3  & -    & 3-kpc arm \\
        & $-$30.1  & 0.8$\pm$0.1  & -\\
        & +26.6  & 1.4$\pm$0.1  & -\\
        & +47.2  & 1.1$\pm$0.1  & +50 km s$^{-1}$ \\ & & & molecular cloud\\
        \midrule
        \multirow{3}{*}{D}  
        & $-$53.3  & 3.2$\pm$0.4  & 3-kpc arm \\
        & $-$32.7  & 1.2$\pm$0.1  & -\\
        & $-$25.0  & 0.5$\pm$0.1  & -\\
        \midrule
        \multirow{4}{*}{E}  
        & $-$55.9  & 1.0$\pm$0.1  & 3-kpc arm \\
        & $-$48.2  & 0.6$\pm$0.1  & -\\
        & $-$25.0  & 0.2$\pm$0.1  & -\\
        & +18.8  & 1.3$\pm$0.1  & -\\
        \bottomrule
    \end{tabular}
    }
    \label{tab:absorption information}
\end{table}

The HI column density and optical depth are related by the formula: $$ N_{HI}=1.8\times10^{18} \times T_s \times \int \tau dv $$ where $N_{HI}$ is the column density of atomic hydrogen in $\text{cm}^{-2}$, $T_s$ is the spin temperature in K, $\tau$ is the optical depth and $v$ is the velocity in km s$^{-1}$. Here, we assumed that the spin temperature does not vary across the absorption line. The HI column density values for two different velocity ranges are shown in Table~\ref{tab:column_density}. These velocity ranges correspond to absorption due to the 3 kpc arm and the +50 km s$^{-1}$ molecular cloud, respectively.

\begin{table}[h]
    \centering
    \caption{Column density values for different parts}
    \begin{tabular}{ccc}
        \toprule
        \multirow{2}{*}{\textbf{Parts}} & \multicolumn{2}{c}{\textbf{Column Density, $N_{HI}$ ($\times 10^{20} \times T_s$)}} \\
        \cmidrule(lr){2-3}
        & $-70$ to $-42$ km s$^{-1}$ & $+33$ to $+90$ km s$^{-1}$ \\
        \midrule
        A & 2.0 & 0.6 \\
        B & 1.1 & 2.7 \\
        C & - & 3.6 \\
        D & 2.1 & - \\
        E & 0.9 & - \\
        \bottomrule
    \end{tabular}
    \label{tab:column_density}
\end{table}

For comparison, we also show the CO emission spectrum \citep{Tokuyama2019} and the CS emission spectrum \citep{Tsuboi1999} towards the same regions in Figure~\ref{fig:hi_co_cs}. CO molecular line emission at around $-53$ km s$^{-1}$ is clearly seen towards all the 5 regions of the 7$^\prime$ halo. Though, there is no clear peak emission seen between $-$25 to $-$30 km s$^{-1}$, but low level of emissions could be seen at those velocities. Strong wide emission feature is also seen towards all the above positions at peak velocities that vary from 53 (regions B, C, D) to 68 (region A) km s$^{-1}$.

An HI absorption study of the central $\sim 100$ pc of the GC was performed by \citet{Lasenby1989} with a spectral resolution of 10 km s$^{-1}$ and an angular resolution of $\sim 1^\prime$. Figure 1 of \citet{Lasenby1989} does show absorption at velocities around $-53$, and +50 km s$^{-1}$, and is, therefore, broadly consistent with our results.  

\section{Discussions}
\label{sec:discussions}
As shown in Figure~\ref{fig:hi_co_cs} and Table~\ref{tab:absorption information}, HI absorption and molecular cloud emission around $-53$ km s$^{-1}$ is clearly seen toward all the 5 parts (regions A to E), as well as in the integrated spectrum (Figure~\ref{fig: HI absorption of 7-arcmin halo}). This emission feature is designated as the 3-kpc arm of the Milky Way \citep{Rougoor1964} which is in the direction of GC and $\sim5$ kpc away from us. Therefore, the absorption at this velocity clearly shows that the 7$^\prime$ halo is located beyond 5 kpc from us.

\begin{figure}
    \centering
    \includegraphics[width=1\linewidth]{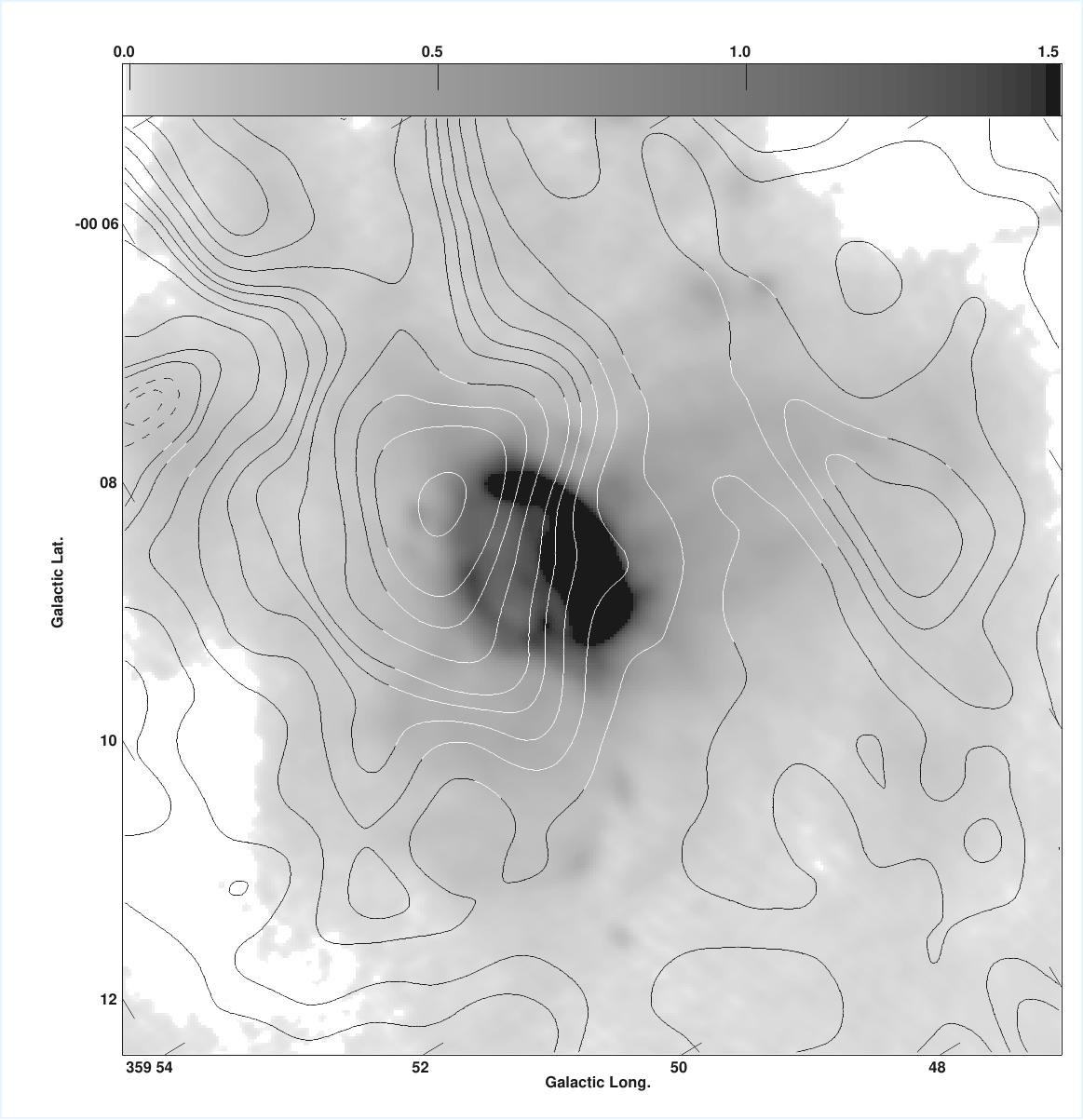}
    \caption{CS molecular emission in contour \citep{Tsuboi1999} with velocity between 40 and 70 km s$^{-1}$ superimposed on a grey scale 1.4 GHz continuum image of 7$^\prime$ halo.}
    \label{fig:cont_image_CS_contour_40_70}
\end{figure}

HI absorption and molecular cloud emission around +50 km s$^{-1}$ are due to the well-known +50 km s$^{-1}$ cloud associated with the Sgr A region (see \citet{Morris1996} for a review). HI absorption from gas associated with the regions marked A, B \& C of the 7$^\prime$ halo indicates at least these parts are behind this cloud. As this cloud is known to be associated with the Sgr A region, this clearly shows the 7$^\prime$ halo to be at the GC or behind it. We have not detected any HI absorption at this velocity towards parts D \& E. In Figure~\ref{fig:cont_image_CS_contour_40_70}, we have shown the CS molecular emission (contour) with velocity between 40 and 70 km s$^{-1}$ superimposed on a grey scale 1.4 GHz continuum image of 7$^\prime$ halo. Figure~\ref{fig:hi_co_cs} \& ~\ref{fig:cont_image_CS_contour_40_70} shows there is molecular emission present around +50 km s$^{-1}$ towards all the parts (A to E) of 7$^\prime$ halo. Therefore, the lack of absorption by HI gas associated with this cloud towards the parts D \& E indicates these regions (D \& E) could be in front of the +50 km s$^{-1}$ molecular cloud and therefore are closer to us. Some parts of the 7$^\prime$ halo (regions A, B, C) are behind and some parts of it (regions D \& E) being in front of the +50 km s$^{-1}$ GC cloud then places the 7$^\prime$ halo at the distance of Sgr A complex. The angular size of 7$^\prime$ halo corresponds to $\sim20$ pc size scale. Parts of it not showing in absorption show that the halo could extend several tens of pcs along our line-of-sight.

CS emission spectrum shown in Figure~\ref{fig:hi_co_cs} from the parts of the 7$^\prime$ halo does not show any peak in emission for all the regions between $-$25 to $-$30 km s$^{-1}$. CS molecular emission is a good tracer of dense molecular clouds. Therefore, the absence of peak CS emission could indicate that the HI absorption seen within the above velocity range is not caused by HI associated with molecular clouds near the GC. It could have been caused by a line-of-sight HI gas in front of the GC region.

By assuming the propagation/expansion velocity of the plasma to be the same as magnetoionic instability, one can estimate its age. Considering a magnetic field of $\sim$1 mG (e.g., \citet{AkshayaandHoang2024}) near the GC, and a plasma density of $\sim$10 cm$^{-3}$ \citep{Cordes2004}, the Alfvén velocity is about 700 km s$^{-1}$. Given the size of the halo is $\sim$20 pc, the corresponding age is $\sim10^5$ years. We also determine the following properties by assuming equipartition of energy between magnetic fields and cosmic ray particles. We estimate the luminosity of the 7$^\prime$ halo over the radio band (10 MHz to 10 GHz) using its spectral index and flux density measured at 10 GHz. Due to low-frequency absorption by foreground ionized gas, the spectral index was determined using the 7$^\prime$ halo's spectrum above 1.4 GHz. The flux densities at wavelengths of 3 cm, 6 cm, and 20 cm are taken from Table 2 of \citet{pedlar1989sgrA}. A least-squares fit to these data, assuming a power-law spectrum of the form $S_\nu = S_0 \nu^{\alpha}$, yields a spectral index of $\alpha \approx -0.6$ and a flux density of 74 Jy at 10 GHz. We derive a total radio luminosity of $L \approx 1.43 \times 10^{35}$ erg s$^{-1}$. The particle energy is $U_p \approx 2.56 \times 10^{50}$ erg, and the corresponding equipartition magnetic field strength is $B(U_\mathrm{min}) \approx 0.2$ mG \citep{Moffet-synchrotron}. This equipartition magnetic field value is comparable with the minimum magnetic field strengths toward the GC threads G0.08+0.15 (0.1 mG) and G359.96+0.09 (70$\mu$G), as reported by \citet{Anantharamaiah1991}. If the GC magnetic field strength is indeed close to its equipartition value, the corresponding age of this halo is $\sim5\times 10^5$ years. Considering the efficiency of conversion of mechanical energy to relativistic particles to be of $\sim10\%$ in a supernova \citep{Energy-Conversion-supernova-review}, it could have been created from the explosions of a few supernovae near the GC (starburst activity). Although the total energy of the relativistic particles in the 7$^\prime$ halo is low, the particle energy density in the halo is $\sim 10^{-9} $ erg cm$^{-3}$, considerably higher than that observed in the lobes of radio galaxies such as Fornax A \citep{Moffet-synchrotron}, and M87 \citep{M87-lofar-equipartition-energy}, where typical values based on the minimum energy principle are $\sim 10^{-12} $ erg cm$^{-3}$. Therefore, the past activity of Sgr A$^*$ could also be responsible for its formation. We note that large-scale bubbles of emission in X-rays and gamma rays have been seen $\sim$10 kpc away from GC, north and south of the Galactic plane. These have been interpreted as the result of the previous activity of Sgr A$^*$ \citep{Fermi-bubble-xray-nature-predehl2020}. These features are believed to be due to the long time scale enhanced activity of Sgr A$^*$. Similarly, its shorter-time-scale enhanced activity could have resulted in the creation of the 7$^\prime$ halo $\sim10^5$ years ago.

\section{Conclusions}
\label{sec:conclusion}
HI absorption study of the 7$^\prime$ halo leads to the following conclusions: 
\begin{enumerate}
    \item Absorption is seen at around $-53$ km s$^{-1}$ due to the 3 kpc arm towards all the 5 regions of the 7$^\prime$ halo, which indicates a minimum distance of $\sim 5$ kpc from us. 
    \item Absorption towards 3 out of 5 parts due to neutral hydrogen associated with the +50 km s$^{-1}$ molecular cloud suggests that the entire 7$^\prime$ halo region might not be at the same distance from us; rather, a part of it to be closer to us (by few tens of pcs).
    \item For the first time, this indicates that it is at the distance of the +50 km s$^{-1}$ molecular cloud, which is believed to be associated with the Sgr A region.
    \item 7$^\prime$ halo could have been created due to energetic activities in the GC region $\sim10^5$ years ago.   
\end{enumerate}

\textit{Acknowledgments:} We acknowledge the support of the Department of Atomic Energy, Government of India, under project No. 12-R\&D-TFR-5.02-0700. We would like to thank the anonymous reviewer for the useful suggestion.

\bibliography{references.bib}
\end{document}